\documentclass[epj]{svjour}
\usepackage{epsfig}
\usepackage{amsmath}
\usepackage{hyperref}
\usepackage{graphics}

\def\eeq{\end{equation}}

\def\prb{Phys. Rev. {B }}

\def\prl{Phys. Rev. Lett. }

\def\ssc{Solid State Commun.}

\def\epl{Euro. Phys. Lett.}

\def\apl{Appl. Phys. Lett.}
\def\jap{J. Appl. Phys.}
\begin{document}
\title{Detecting a true quantum pump effect}

\author{Colin  Benjamin}

\institute{ Centre de Physique Theorique, CNRS.UMR 6207-Case 907,
Faculte des Sciences de Luminy, 13288 Marseille Cedex 09, France.}
\date{}
\abstract{ Even though quantum pumping is a very promising field,
it has unfortunately not been unambiguously experimentally
detected. The reason being that in the experiments the
rectification effect overshadows the pumped current. One of the
better known ways to detect it is by using the magnetic field
symmetry properties of the rectified and pumped currents. The
rectified currents are symmetric with respect to magnetic field
reversal while the pumped currents do not possess any definite
symmetry with respect to field reversal. This feature has been
exploited in some recent works. In this work we look beyond this
magnetic field symmetry properties and provide examples wherein
the nature or magnitudes of the pumped and rectified currents are
exactly opposite enabling an effective distinction between the
two.}


\maketitle

\section{Introduction}

Quantum pumping is an unique way to transport charge or spin
without applying any voltage bias\cite{glazman_science,lubkin}.
The idea of quantum pumping has been around for a long time
beginning with the works of Thouless in Ref.\cite{thouless} and
Niu in Ref.\cite{niu} and later with the works of Buttiker, Thomas
and Pretre in Ref.\cite{bpt}, Brouwer in Ref.\cite{brou_pump} and
Zhou, Spivak and Altshuler in Ref.\cite{zsa}. Regrettably the
unambiguous detection of this effect has not been possible till
date\cite{mosk_but,brou_rect}. The experiment\cite{marcus_sci}
which was originally thought to be a quantum pumping experiment is
now universally accepted as a detection of rectified
currents\cite{marcus_recent}. Although there might have been a
pumped current which unfortunately was masked by the rectified
currents\cite{mosk_but,brou_rect,mucciolo}. Experimentally, what
seems to happen in pumping experiments is that the time dependent
parameters may through stray capacitances directly link up with
the reservoirs and thus indirectly induce a bias which is the
origin of the rectified current\cite{brou_rect}. The reason why
the urgent detection of a true quantum pump effect is immediately
required is because manifold theoretical proposals based on
quantum pumping ranging from the use of the quantum pump effect to
drive a pure spin current\cite{ronald,qspin} to the use of quantum
pump effect as a means for quantum computation\cite{buti_been}
have come up. With so much at stake an early resolution of this
vexed question is not only necessary but also urgently required.
This work proposes to answer this question.

 Now how to detect pumped and rectified currents if both are
present in a single experiment. One of the ways is to look at the
symmetries with respect to magnetic field reversal these currents
possess\cite{mosk_but,shutenko,kamenev}. To further explain the
preceding statement let us start from the definitions of  the
rectified and pumped currents. The rectified current in a two
terminal setup is given by\cite{brou_rect}:
\begin{equation}
I_{rect}=\frac{w}{2\pi}R\int_{S}dX_{1}dX_{2}(C_{1} \frac{\partial
G}{\partial X_1}- C_{2} \frac{\partial G}{\partial X_2})
\end{equation}
Herein $R$ is the resistance of circuit path and is assumed to be
much less than the resistance of the mesoscopic scatterer, while
$C_1$ and $C_2$ are stray capacitances which link the gates to the
reservoirs, $X_1$ and $X_2$ are the modulated gate voltages.
Finally, $G$ is the Landauer conductance which is just the
transmission probability (T) of the mesoscopic scatterer. The
pumped current into a specific lead in a two terminal system, is
in contrast given as\cite{brou_pump}
\begin{equation}
I_{pump}=\frac{e}{\pi}\int_{A}dX_{1}dX_{2}\sum_{\beta}\sum_{\alpha
\in 1} Im(\frac{\partial S_{\alpha \beta}^{*}}{dX_1}\frac{\partial
S_{\alpha \beta}^{}}{dX_2})
\end{equation}

In the above equation, $S_{\alpha\beta}$ defines the scattering
amplitude (reflection/transmission) of the mesoscopic sample, the
periodic variation of the parameters $X_1$ and $X_2$ follows a
closed path in a parameter space and the pumped current depends on
the enclosed area $A$ in ($X_1$,$X_2$) parameter space. Initially,
the mesoscopic sample is in equilibrium and for it to transport
current one needs to simultaneously vary two system parameters
$X_{1}(t)=X_{1}+\delta X_{1} \sin (wt)$ and $X_{2}(t)=X_{2}+\delta
X_{2} \sin (wt+\phi)$, herein $\delta X_{i}$ defines the amplitude
of oscillation of the adiabatically modulated parameters. In the
adiabatic  quantum pumping regime we consider the system thus is
close to equilibrium\cite{mosk_ref}.

 The essential difference between the rectified currents and
the pumped currents are while the former is bound to be symmetric
with respect to magnetic field reversal (via, Onsager's symmetry)
since the conductance\cite{buttiker} and it's derivatives enter
the formula, the pumped currents would have no definite symmetry
with respect to magnetic field reversal\cite{shutenko,kamenev}
since they in turn depend on the complex scattering amplitudes
which have no specific dependence on field reversal unless the
scattering system possesses some specific discrete
symmetries\cite{kamenev}. The main motivation of this work is to
provide examples beyond the distinctive properties the two
currents possess with respect to magnetic field reversal.

 The examples show that the currents can be easily differentiated,
either there natures are so different or their magnitudes are so
very different that it enables an easy detection. The three
examples provided are: (1) pumped and rectified currents in
presence  of magnetic barriers, (2) pumped and rectified currents
in a normal metal double barrier structure and finally (3) pumped
and rectified currents at a normal metal- charge density wave
interface. In example (1) while the pumped currents are cent
percent spin polarized the rectified currents are completely
unpolarized, in example (2) pumped current is finite while the
rectified current is zero, and finally in example (3) the
rectified current again is zero while pumped current is finite. Of
course these examples are by no means the only examples that can
be found there might be numerous other examples wherein the pumped
and rectified currents vary in such a distinct fashion apart from
of-course the distinction brought out by magnetic field symmetry.
In Ref.\cite{fazio} the authors consider a three terminal
structure with a single normal metal lead with two superconducting
leads. The pumped current into the normal metal lead has no
definite symmetry with respect to the phase of the order parameter
while the conductance is symmetric in phase. In another
interesting work\cite{mares}, the effect of dephasing was
considered and it was shown that effect of dephasing on
rectification effects is less pronounced than for quantum pumping.

 The rectified currents in the adiabatic quantum pumping regime
we consider differ from that in the non-linear dc bias regime. In
the latter the Onsager symmetry relations are not
obeyed\cite{non-linear} while in the former (from Eq. 1) they are
obeyed. Further rectification can also be talked of when a high
frequency electromagnetic field is applied to a phase coherent
conductor\cite{falko}. This case also falls into the non-linear
regime.

Our motivation in this work is plain. We provide three examples
wherein the distinctive characteristics of the pumped  and
rectified currents are brought out. The symmetry properties these
currents have with respect to magnetic field reversal are not as
clear cut as it would seem initially. For example in
Ref.\cite{kamenev} it was pointed out that if the mesoscopic
scatterer has some distinct spatial symmetries then the pumped
current itself can be symmetric with respect to magnetic field
reversal. Our work hopefully will provide a compass which would
point into clear blue water between rectified and pumped currents.

\section {Examples}
\label{sec:theory} In the examples below we look into the weak
pumping regime for both the rectified as well as the pumped
currents, since we can derive analytical expressions in this
regime. The weak pumping regime is defined as one wherein the
amplitude of modulation of the parameters is small, i.e., $\delta
X_{i} \ll X_{i}$. In the weak pumping regime the rectified current
reduces to:

\begin{equation}
I_{rect}=I_{rect}^{x} [C_{1}\frac{\partial T}{\partial
X_{1}}-C_{2}\frac{\partial T}{\partial X_{2}}]
\end{equation}
with $I_{rect}^{x}=we^{2}\sin(\phi)\delta X_{1}\delta X_{2}
R/4\pi^{2}\hbar$. $T$ is the transmission  coefficient of the
mesoscopic scatterer, and for the special case of capacitances
with equal magnitude, i.e., $C_{1}=C_{2}=C$ one has:
\begin{equation}
I_{rect}=I_{rect}^{0} [\frac{\partial T}{\partial X_{1}}-
\frac{\partial T}{\partial X_{2}}]
\end{equation}
with $I_{rect}^{0}=we^{2}\sin(\phi)\delta X_{1}\delta X_{2}
RC/4\pi^{2}\hbar$.
 Similarly the pumped current into lead $\alpha$ are:
\begin{equation}
I_{pump,\alpha}=I_{pump}^{0}\sum_{\beta}\sum_{\alpha \in 1}
Im(\frac{\partial S_{\alpha \beta}^{*}}{dX_1}\frac{\partial
S_{\alpha \beta}^{}}{dX_2})
\end{equation}
with $I_{pump}^{0}=we \sin(\phi)\delta X_{1}\delta X_{2}/2\pi$,
$w$ is the frequency of the applied time dependent parameter,
$\phi$ is the phase difference between the parameters and $e$ is
the electronic charge.

\subsection{Magnetic barrier's}

The first example is of pumping and rectification in case of a
magnetic barriers. The model of our proposed device is exhibited
in Fig.~1. It is essentially a 2DEG in the $xy$ plane with a
magnetic field in the z-direction. The magnetic field profile we
consider is of delta function type for simplicity, ${\bf
B}=B_{z}(x)\hat z$ with $B_{z}(x) =
B_{0}[\delta(x+d/2)-\delta(x-d/2)]$, wherein $B_0$ gives the
strength of the magnetic field and $d$ is the separation between
the two $\delta$ functions (see Fig.1(c)). The above form of the
magnetic field is an approximation of the more general form seen
when parallely magnetized ferromagnetic materials are
lithographically patterned on a 2DEG (Fig.1(b)). This
approximation is not novel to this work but has been used in a
number of works, see Ref.[21] for further details. Magnetic
barrier's can not only be formed by this method but also when a
conduction stripe with current driven through it is deposited on a
2DEG, and also when a super-conductor plate is deposited on a
2DEG, see Refs.\cite{mag_bar_peeter,mag_bar_lu} for details. The
structure depicted in Figure 1 has been experimentally produced as
shown in Ref.\cite{mag_bar_peeter}. There are a host of
experiments\cite{mag_expt} wherein such type of and similar
structures are made, discussed and transport measurements carried
out.

A 2DEG in the $xy$ plane with a magnetic field pointing in the $z$
direction is described by the Hamiltonian-

\begin{eqnarray}
H&=&\frac{1}{2m^{*}} [{\bf p} +e{\bf A}(x)]^{2} +
\frac{eg^{*}}{2m_{0}}
\frac{\sigma \hbar}{2} B_{z}(x)  \nonumber\\
&=&\frac{1}{2m^{*}} ({p_{x}}^{2}+[p_{y}+eA(x)]^{2})+
\frac{eg^{*}}{2m_{0}}\frac{\sigma
  \hbar}{2} B_{z}(x)
\end{eqnarray}

where $m^{*}$ is the effective mass of the electron, $\bf {p}$ is
it's momentum, $g^{*}$ the effective g-factor and $m_{0}$ is the
free-electron mass in vacuum , $\sigma=+1/-1$ for up/down spin
electrons, and ${\bf A}(x)$, the magnetic vector potential is
given in the landau gauge for the region $-d/2< x < d/2$ and for
incoming electrons from the left by ${\bf A}(x)=B_{0}\hat y$, and
for electrons incoming from the right by ${\bf A}(x)=-B_{0}\hat
y$, The magnetic vector potential is zero otherwise. The last term
in Eq.~6 is zero everywhere except at $x=\pm d/2$. For simplicity
we introduce dimensionless units, the electron cyclotron frequency
$w_{c}=eB_{0}/m^{*}c$, and the magnetic length $l_{B}=\sqrt{\hbar
  c/eB_{0}}, with B_{0} $ being some typical magnetic field. All the
quantities are expressed in dimensionless units: the magnetic
field $B_{z}(x)\rightarrow B_{0} B_{z}(x)$, the magnetic vector
potential ${\bf{A}(x)}\rightarrow B_{0}l_{B}\bf{A}(x) $, the
coordinate ${\bf{x}}\rightarrow l_{b}x$ and the energy
$E\rightarrow \hbar w_{c} E(=E_{0}E)$.

Since the Hamiltonian as depicted in Eq.~6 is translation-ally
invariant along the y-direction, the total wave-function can be
written as $\Psi(x,y)=e^{i q y} \psi(x)$, wherein $q$ is the
wave-vector component in the y-direction. Thus one obtains the
effective one-dimensional Schroedinger equation-

\begin{figure}
  \protect\centerline{\epsfxsize=3.0in\epsfbox{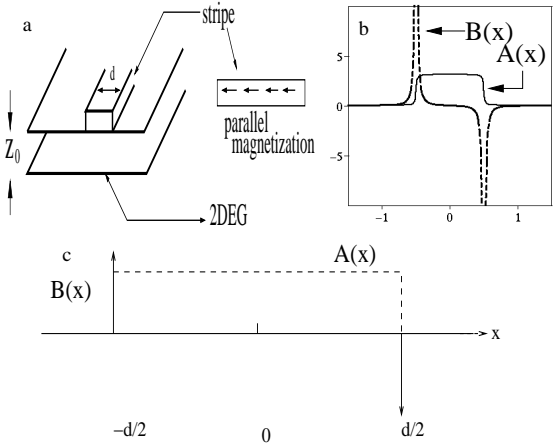}}
\caption{(a) The device- On top of a 2DEG a parallely magnetized
  magnetic stripe is placed. (b) The realistic magnetic field profile
  in a 2DEG along-with the magnetic vector potential for the device
  represented in (a). (c) The model magnetic field (delta function
  B(x)) profile along with the magnetic vector potential A(x).}
\end{figure}
\begin{equation}
[\frac{d^{2}}{dx^{2}}-\{A(x)+q\}^{2}-
\frac{eg^{*}}{2m_{0}}\frac{\sigma
  m^{*}}{\hbar} {B_{z}(x)} + \frac{2 m^{*}}{\hbar^{2}} E] \psi(x)=0
\end{equation}

The S-matrix for electron transport across the device can be
readily found out by matching the wave functions and as there are
$\delta$ function potentials there is a discontinuity in the first
derivative. The wave functions on the left and right are given by
$\psi_1 =(e^{ik_{1}x}+r e^{-ik_{1}x})$ and $\psi _3=t
e^{ik_{1}x}$, while that in the region $-d/2<x<d/2$ is $\psi_2=(a
e^{ik_{2}x}+b e^{-ik_{2}x})$. The wave vectors are given by-
$k_{1}=\sqrt{2E-q^2}$, $k_{2}=\sqrt{2E-(q+B_{z})^2}$ and for
electrons incident from the right, $k_2$ in the wave-functions is
replaced by $k_{2}^{\prime}=\sqrt{2E-(q-B_{z})^2}$. With this
procedure outlined above one can determine all the coefficients of
the S-Matrix

\[S_{\sigma}=\left(\begin{array}{cc}
s_{\sigma 11}       & s_{\sigma 12} \\
s_{\sigma 21} &     s_{\sigma 22}  \\
\end{array} \right)=\left(\begin{array}{cc}
r_{\sigma}       & t^{\prime}_{\sigma} \\
t_{\sigma} &     r^{\prime}_{\sigma}  \\
\end{array} \right) \]

\begin{eqnarray*}
r_{\sigma}&=&\frac{-i \sin(k_{2} d) (k^{2}_{1}-k^{2}_{2}-\lambda^{2}-2i\lambda\sigma k_{1})}{D} \\
t_{\sigma}&=&\frac{2 k_{1} k_{2}}{D}, t^{\prime}_{\sigma}=\frac{2 k_{1} k'_{2}}{D'}\\
r^{\prime}_{\sigma}&=&\frac{-i \sin(k'_{2} d) (k^{2}_{1}-k'^{2}_{2}-\lambda^{2}+2i\lambda\sigma k_{1})}{D'} \\
\mbox{ with }
D&=&2 k_{1} k_{2} \cos(k_{2} d) - i \sin (k_{2} d)(k^{2}_{1} +k^{2}_{2} +\lambda^{2}), \\
D'&=&2 k_{1} k'_{2} \cos(k'_{2} d) - i \sin (k'_{2} d)(k^{2}_{1} +k'^{2}_{2} +\lambda^{2}), \\
\lambda&=&\frac{g^{*} B_z}{2},\mbox{  }  k_{1}=\sqrt{2E},
k_{2}=\sqrt{2E-(q+B_{z})^{2}}\\
\mbox{and }k'_{2}&=&\sqrt{2E-(q-B_{z})^{2}}.
\end{eqnarray*}

One can readily see from the transmission coefficients, there is
no spin polarization as $T_{+1}=T_{-1}$. This type of structure
has already been studied in Ref.\cite{ronald} where it's
remarkable pure pumped spin current properties were noticed. In
this work we compare and contrast the pumped currents with the
rectified currents and show that the rectified currents are
completely unpolarized. This provides an unique way to distinguish
the two effects. The schematic of the system is exhibited in Fig.
1. We in the following consider $q=0$, and therefore
$k^{\prime}_{2}=k_{2}$.

Initially, the device is in equilibrium, and for it to transport
current one needs to simultaneously vary two system parameters
$X_{1}(t)=X_{1}+\partial X_{1}\sin(wt)$ and
$X_{2}(t)=X_{2}+\partial X_{2}\sin(wt+\phi)$, in our case $X_{1}$
is the width $d$ and $X_{2}$ the magnetic field $B_{z}$ given in
terms of the magnetization strength $B_{0}=M_{0}h$, where $h$ is
the height and $M_0$ the magnetization of the ferromagnetic
stripe. To invoke pumping in our proposed system we modulate the
width ($d=d_{0}+d_{p}\sin(wt)$) and magnetic field strength
($B_{z}=B_{x}+B_{p}sin(wt+\phi)$). Herein $w$ is the pumping
frequency and $\phi$ is the phase difference between the two
modulated parameters. Thus in this adiabatic pumping regime the
system is close to equilibrium.
\begin{figure}
\vskip 0.7cm
  \protect\centerline{\epsfxsize=3.5in\epsfbox{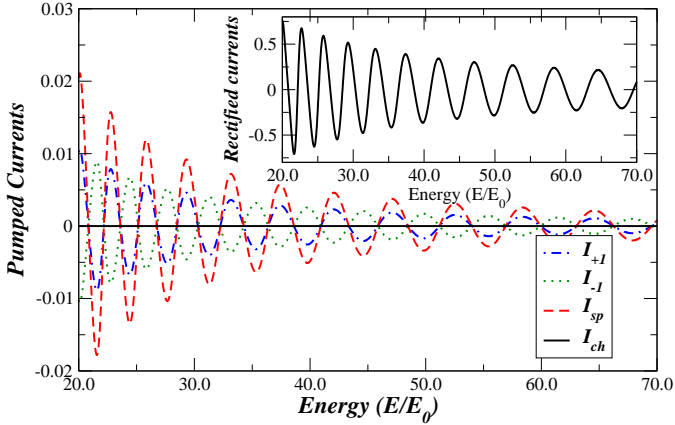}}
\caption{(Color online) Energy dependence of the pumped current
normalized by $I^{0}_{pump}$. Spin polarized
  pumping delivering a finite net spin current along-with zero
  charge current.  The parameters are $B_{x}=5.0, d_{0}=5.0,
  \phi=\pi/2, g^{*}=0.44$ and wave-vector $q=0$. In the inset the
  rectified currents are plotted. The rectified currents normalized by
$I^{0}_{rect}$ (for same parameters as for pumping) are
  of-course completely unpolarized. }
\end{figure}
 The transmission coefficient of this structure which in
effect is the Landauer conductance is-
\begin{eqnarray}
T&=&\frac{4k_{1}^{2}k_{2}^{2}}{4k_{1}^{2}k_{2}^{2}\cos^{2}(k_{2}d)+(\lambda^{2}+k_{1}^{2}+k_{2}^{2})\sin^{2}(k_{2}d)}\nonumber\\
& &\mbox{with, } k_{1}=\sqrt{2E}, k_{2}=\sqrt{2E-B_{z}^{2}}
\mbox{and } \lambda_{1}=\frac{g^{*}B_{z}}{2}.
\end{eqnarray}

As is self evident, the transmission is completely unpolarized,
i.e. $T_{\sigma}=T_{-\sigma}$. This fact was discovered only in
Ref.\cite{papp}, two earlier works\cite{amlan} had mistakenly
attributed spin polarizability properties to the device (as
depicted in Fig. 1) when a bias is applied. Further because of the
fact that spin polarization is absent in presence of a bias, there
wont be any spin accumulation\cite{levy} either. Hence from Eq. 4,
since the rectified current involves the derivatives of the
conductance with respect to the modulated parameters as in Eq. 4,
$ X_{1}= B_z$ and $ X_{2}= d$, the rectified current is completely
unpolarized.  The explicit expression for the rectified and pumped
currents are:

\begin{eqnarray*}
I_{rect}&=& I^{0}_{rect}\frac{-\sin^{2}(k_{2}d)
f'+k_{2}\sin(2k_{2}d)(f-1)}{T'^{2}_d},\nonumber\\
 \mbox{with }
f&=&[\frac{\lambda_{1}^2 +
k_{1}^2+k_{2}^2}{2k_{1}k_{2}}]^2,\nonumber\\
f'&=&\frac{(2EB_{z}(4+g^{2}B)+B_{z}^3(g^{2}-4)(1-B_{z}))}{(64E(2E-B_{z}^{2})^{2})}\nonumber\\
& &\mbox{and } T'_{d}=\cos^{2}(k_{2}d)
+f\sin^{2}(k_{2}d).\nonumber
\end{eqnarray*}
In contrast the pumped currents as in Ref.\cite{ronald}, are given
as:
\begin{eqnarray*}
I_{\sigma}&=&\sigma I^{0}_{pump}\frac{2B^{2}_{z}g^{*}g^{\prime}k^{3}_{1}k^{2}_2\sin(2k_{2}d)}{T^{2}_d},\nonumber\\
I_{sp}&=&I_{+1}-I_{-1}=I^{0}_{pump}\frac{4B^{2}_{z}g^{*}g^{\prime}k^{3}_{1}k^{2}_2\sin(2k_{2}d)}{T^{2}_d},\nonumber\\
I_{ch}&=&I_{+1}+I_{-1}=0,\nonumber
\end{eqnarray*}
\begin{eqnarray*}
& &\mbox{with }g^{\prime}=1-\frac{g^{*2}}{4},\nonumber\\
T_{d}&=&4k^{2}_{1}k^{2}_{2}\cos^{2}(k_{2}d)+[4E-g^{\prime}B^{2}_{z}]^{}\sin^{2}(k_{2}d),\nonumber\\
& &\mbox{} I^{0}_{pump}= \frac{ewB_{p}d_{p}\sin(\phi)}{2\pi},
\mbox{and }\nonumber\\
I_{rect}^{0}&=&we^{2}\sin(\phi)B_{p}d_{p}RC/4\pi^{2}\hbar.
\nonumber
\end{eqnarray*}

The rectified currents as is evident from the above equations are
completely unpolarized, while the pumped currents are completely
spin polarized. There is net zero pumped charge current while a
finite pure spin current flows. In Fig.~2 we show the plots for
the pumped currents with the rectified currents plotted in the
inset of the figure. The figure for the rectified currents is for
equivalent coupling of stray capacitances, but the unpolarized
nature of the rectified current will be valid as well in case of
non-equivalent stray capacitances, since the transmission is
completely unpolarized. For $q \neq 0$, as before we have
completely unpolarized rectified currents, but in the pumping
regime we no longer have pure spin pumped polarized currents but
both pumped finite spin and charge currents. Thus the system can
again discriminate between pumped and rectified currents but not
as as effectively as for the $q=0$ case.

To conclude the analysis of magnetic barriers, we have shown
distinct properties of the rectified and pumped currents. The
experimental realization of such type of structures has already
been achieved. The only thing one has to add is to adiabatically
modulate two independent parameters of our structure (to derive
the currents above we have modulated the width of the magnetic
barrier and it's strength) to see the distinctive spin
polarizability properties the currents possess. To do this one can
make a point contact between the ferromagnetic stripe and the 2DEG
interface applying an ac gate voltages to this point contact can
change the shape of the structure while to change the strength of
the barrier one can apply an external time dependent magnetic
field to the ferromagnetic stripe.

\subsection{Normal metal double barrier structure}
\begin{figure}[h]
\protect\centerline{\epsfxsize=3.0in\epsfbox{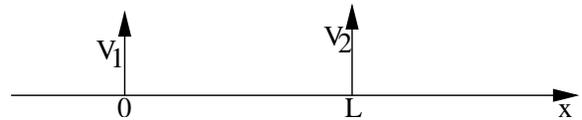}}
\caption{The double barrier structure.The normal metal double
barrier structure is defined via the potential:
$V_{1}\delta(x)+V_{2}\delta(x-L)$.}
\end{figure}

In these type of structures pumping has again been studied as in
Ref. \cite{wang_dbrs}. We consider two $\delta$ function
potentials separated by a length $l$ as in Fig. 3. The
transmission and reflection amplitudes for such type of structures
can be easily calculated by matching the wave-functions at the
three interfaces and then by taking into account the jump in the
first derivative at the interfaces. The transmission coefficient
for this structure is given as:
\begin{eqnarray}
T&=&\frac{4}{a^{2}+b^{2}}\\
\mbox{with, }
a&=&z_{1}z_{2}\sin(kl)+(z_{1}+z_{2})\cos(kl)-2\sin(kl),\nonumber\\
\mbox{} b&=&2 \cos(kl)+(z_{1}+z_{2})sin(kl),\nonumber \mbox { and
$z_{i}=\frac{mV_{i}}{\hbar^{2}k}$}.
\end{eqnarray}

\begin{figure}
\vskip 0.5cm
\protect\centerline{\epsfxsize=3.0in\epsfbox{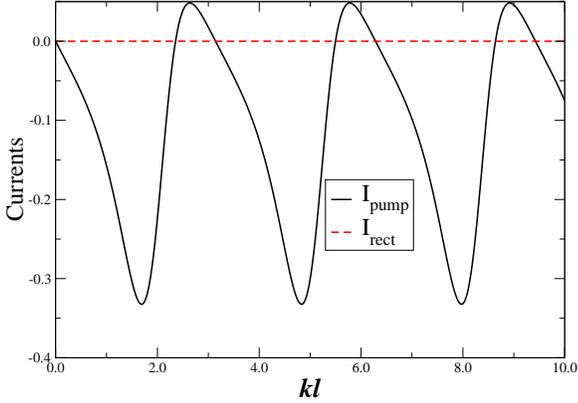}}
\caption{(Color online) The pumped (normalized by $I^{0}_{pump}$)
and rectified currents (normalized by $I^{0}_{rect}$) for a normal
metal double barrier structure as a function of the dimensionless
wavevector $kl$. The strengths of the delta function barriers
$z_{1}=z_{2}=1.0$}
\end{figure}

To invoke pumping in these structures we modulate the strengths of
the barrier potentials, thus $z_{1}=z_{01}+z_{p}\sin(wt)$, and
$z_{2}=z_{02}+z_{p}\sin(wt+\phi)$. One can easily notice from the
Eq. 4, that for barriers of equivalent strength $\frac{\partial
T}{z_{1}}=\frac{\partial T}{z_{2}}$ and the system wont transport
any current but a finite pumped current exists. The rectified
currents are given by from Eq. 4,
\begin{eqnarray}
&&I_{rect}=\frac{-4I^{0}_{rect}}{(a^{2}+b^{2})^{2}}[2z_{1}z_{2}(z_{1}-z_{2})\sin^{2}(kl)\nonumber\\
&+&(z^{2}_{1}-z^{2}_{2})\sin(2kl)+2(z_{1}-z_{2})(\cos(2kl)-1)]
\end{eqnarray}
 The rectified current
thus by Eq. 4 is zero for $z_{1}=z_{2}$.  In Eq. 10,
$I^{0}_{rect}= we^{2}\sin(\phi)z^{2}_{p}RC/4\pi^{2}\hbar$. Of
course one must note that this is in addition for equi-potential
barriers is valid only if the strengths of the stray capacitances
as in Eq. 4, are also equivalent. Further the pumped currents in
the weak pumping regime $z_{p} \ll z_{0i}, i=1,2$, can also be
easily derived (from Eq. 5) and are written below, again
 for $z_{1}=z_{2}=z$:
\begin{equation}
I_{pump}=I^{0}_{pump}\frac{-8\sin(kl)(z\sin(kl)+\cos(kl))}{(a^{2}+b^{2})^{2}}
\end{equation}
Here again $a$ and $b$ are as given in Eq. 10, and  $I^{0}_{pump}=
we^{2}\sin(\phi)z^{2}_{p}/\pi$. Thus for barriers of equal
magnitude the pumped current is finite while rectified currents
are zero. Here we show that the rectified currents are zero in
contrast while pumped currents are finite. Of-course this result
is subject to the condition that the capacitances $C_{1}=C_{2}$.
In Fig. 3, we plot the rectified currents and pumped currents for
such a structure.

The experimental realization of this structure is not at all
difficult, since double barrier structures have been
experimentally realized for long. The only thing is by having two
ac dependent gate voltages to modulate the shape of the double
barrier structure such that the coupling to the stray capacitances
may be equal. If this condition is realized then this very simple
structure will be a very good identifier of a genuine quantum pump
effect if present. Of-course not any structure with equivalent
stray capacitances will give zero rectified current nor would any
device with equi-potential barriers, the most important fact is
the equality $dT/dz_{1}=dT/dz_{2}$, which has to satisfied for the
absence of rectified currents.

\subsection{Normal metal- Charge density wave interface}

Finally we show that pumping and rectification currents at a
normal metal charge density wave interface can also be easily
distinguished since the pumped currents are finite while rectified
currents are again zero. Since the conductance is effectively zero
this result is in fact independent of whether or not
$C_{1}=C_{2}$. We consider a normal metal - charge density wave
junction with an interface at $x=0$ as in Fig. 5. In the charge
density wave region ($x > 0$) the order parameter
$\Delta(x)=\Delta e^{i \chi}$ near the interface is not constant
but decays smoothly over a finite length of the order of the
coherence length\cite{visscher_prb}. This is the charge density
wave proximity effect. In our analysis of the problem we disregard
the proximity effect and assume a step function pair potential.
The structure we work with is depicted in Fig. 5.
\begin{figure}[h]
\protect\centerline{\epsfxsize=3.0in\epsfbox{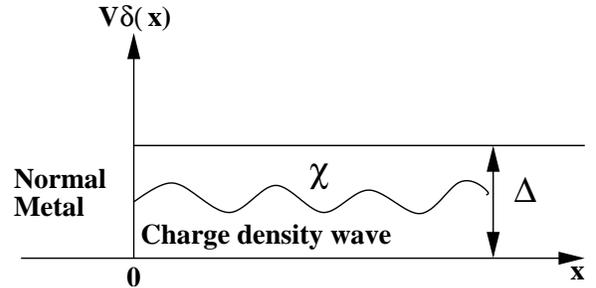}}
\caption{The normal metal-charge density wave interface. We
disregard the proximity effect. $\Delta$ denotes the strength of
the order parameter of the charge density wave while $\chi$
denotes its phase.}
\end{figure}
\begin{figure}
\vskip 0.5in
  \protect\centerline{\epsfxsize=3.0in\epsfbox{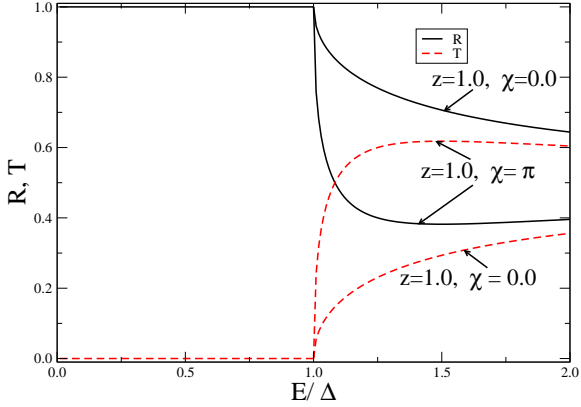}}
\caption{(Color online) The transmission and reflection
probabilities are plotted (parameters are mentioned in the
figure). As is evident the transmission is zero in the tunnelling
regime. }
\end{figure}

A delta function potential $V\delta(x)$at the interface models the
impurity which pins the charge density wave. We also assume the
charge density wave and normal metal to be one dimensional and
average electron densities are equal. The fermi wave-number $k_F$
and the effective masses are assumed to be equal in the normal
metal and charge density wave regions. The scattering matrix of
such a junction has been derived earlier in
Ref.\cite{visscher1,tanaka}. Here we give the results. The
scattering amplitudes of the structure depicted in Fig. 5 are
given below:

\begin{eqnarray}
r&=&\frac{-iz(u+ve^{-i\chi})+ve^{-i\chi}}{(1+iz)u+izve^{-i
\chi}},\\
   r'&=&-\frac{(1+iz)ve^{i \chi}+izu}{(1+iz)u+izve^{-i
\chi}},\nonumber\\
t&=&\sqrt{\frac{k}{q}}\frac{1}{(1+iz)u+izve^{-i \chi}},\\
   t'&=&\sqrt{\frac{q}{k}}\frac{u^{2}-v^{2}}{(1+iz)u+izve^{-i
\chi}}\\
\mbox{with} &&u^{2}=\frac{1}{2}(1+\frac{w}{E}),
v^{2}=\frac{1}{2}(1-\frac{w}{E}),\\
&&\mbox{        in the  propagating regime,}\nonumber\\
\mbox{and}&&u^{2}=\frac{1}{2}(1+\frac{i w'}{E}),
v^{2}=\frac{1}{2}(1-\frac{i w'}{E}),\\
&& \mbox{          in the tunnelling regime. }\nonumber
\end{eqnarray}

In the above expressions, $w=\sqrt{E^{2}-\Delta^{2}}$,
$w'=i\sqrt{\Delta^{2}-E^{2}}$  and $z=V/\hbar v_{F}$, with
$q=E/\hbar v_{F}$ and $k=w/\hbar v_{F}$ in the propagating regime
while $k=iw'/\hbar v_{F}$ in the tunnelling regime, $\chi$ is
phase of the charge density wave.

 The unique thing of such a normal metal-insulator-charge density
wave structure is that the macroscopic phase ($\chi$) of the
charge density wave appears in the expression for the transmission
$|t|^2$ and reflection $|r|^2$ probabilities.  This is in sharp
contrast to a normal metal-insulator-superconductor structure
where the macroscopic phase of the superconductor does not appear
in the transmission and reflection probabilities. Here of course
we are interested in the distinct characteristics of the rectified
current and the pumped current. The unique thing of our structure
is that in the tunnelling regime for $E \ll \Delta^2$, the system
does not conduct (as $|t|^2=0$, see Fig. 6) but pumps a finite
current as in Fig. 7. This is because the transmission probability
is zero which can easily be seen also from the above equation,
while in the same regime there is a finite pumped current. To
invoke pumping in this structure we modulate the strength of the
delta function barrier ($z=z_{0}+z_{p}\sin(wt)$) and the phase of
the charge density wave order parameter
($\chi=\chi_{0}+\chi_{p}\sin(wt+\phi)$).
\begin{figure}
\vskip 0.5cm
\protect\centerline{\epsfxsize=3.0in\epsfbox{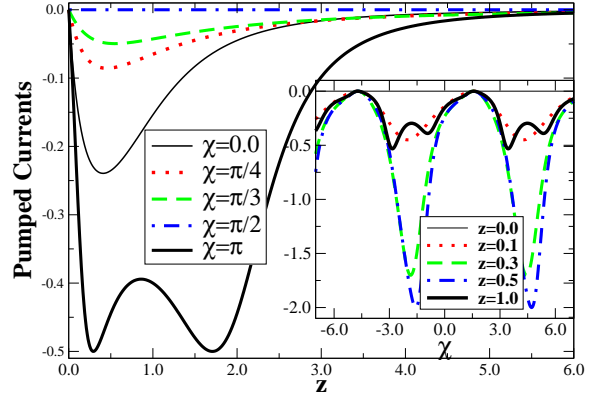}}
\caption{(Color online) The pumped current into the Charge density
wave material at the interface between a normal metal and a charge
density wave interface  is of course finite in the tunnelling
regime ($E \ll \Delta$). In the main panel the pumped currents are
as function of the barrier strength $z$ for different values of
the phase difference $\chi$, while in the inset the currents are
plotted as function of the phase difference $\chi$ for different
values of the barrier strength $z$.}
\end{figure}

 In Fig.~7, we plot the  pumped
currents into the charge density wave material for such a
structure in the tunnelling regime. The transmission and
reflection coefficients are also plotted in Figure 6, which bring
out the fact that there is no transport in the tunnelling regime.
The plot clearly brings out the differences as the rectified
currents in the tunnelling regime are exactly zero while a finite
pumped current exists. The expression for the pumped current can
also be easily derived in the weak pumping regime $z_{p} \ll
z_{0}$ and $\chi_{p} \ll \chi_{0}$ (see Eq. 4), and in tunnelling
regime, i.e., the limit where $E \ll \Delta$-

%%\begin{equation}
%I_{pump}=I^{0}_{pump}\frac{-2\Delta \cos(\chi) +4z\Delta
%\sin(\chi)+4z^{2}\Delta\cos(\chi)-2z\Delta-\sqrt{2\Delta}(z\cos(\chi)-\sin(\chi))}{4\Delta
%z^{2}(\sin(\chi)-1)^{2}[1+2z\cos(\chi)+2z^{2}]+\Delta(1+2z\cos(\chi))[1+2z\cos(\chi)-4z^{2}\sin(\chi)(\sin(\chi)-1)]}
%\end{equation}

\begin{equation}
I^{CDW}_{pump}=\frac{2zI^{0}_{pump}[\sin(\chi)-1]}{a_{z}+b_{z}\cos(\chi)+c_{z}\sin(\chi)-\cos^{2}(\chi)[d_{z}-f_{z}(\chi)]}
\end{equation}
with, $I^{0}_{pump}= we^{2}\sin(\phi)z^{}_{p}\chi_{p}/\pi,
a_{z}=1+8z^{4}, b_{z}=4z(1-2z), c_{z}=4z^{2}(1-2z^{2}),
d_{z}=4z^{2}(z^{2}+2z-3),
f_{z}(\chi)=8z^{3}(\cos(\chi)+\sin(\chi)).$

One can easily see that when the delta function which pins the CDW
is absent, i.e., $z=0$, there is no pumped current. Further when
$\chi=\pi/2$  there is again no pumped current. Apart from these
two cases the system pumps a finite pumped current for all other
values.  Browser's formula as in Eq. 2, was derived for same
particles carrying current at both sides of a scatterer. But
Brouwer's formulation has been generalized to Normal metal
-superconductor junctions\cite{jian}. In normal
metal-superconductor junctions below the energy gap there cannot
be any quasi particle transmission, but there is andreev
reflection which results in cooper pair transport into
superconductor. Something similar happens here below the energy
gap. Here there are no cooper pairs, there are instead
electron-hole pairs further there is no analog of andreev
reflection. What happens when the system is biased is that there
is no quasi particle transport into the CDW, since the
transmission probability is zero. Since the pumped currents are
described by amplitudes reflection and transmission one has a
finite pumped current into the CDW. The pumped current into CDW is
of-course made of electron-hole pairs.

One can also describe the pumped current into normal metal. One
can also distinguish between rectification and pumping via the
currents in the normal metal. There is of-course no net rectified
current transported into the normal metal lead as whatever is
incident at the interface is completed reflected in the tunnelling
regime ($R=1$). In-contrast the pumped current is finite and in
figure 8 we plot the pumped currents into the normal metal lead.
The pumped characteristics can be seen from Eq. 18, for either
$z=0$ or $\chi=\pi/2$ there is no pumped current similar to the
pumped current into the CDW material.
\begin{equation}
I^{N}_{pump}=\frac{2zI^{0}_{pump}[z\sin(2\chi)+\sin(\chi)-1-2z\cos(\chi)]}{4z^{4}a(\chi)-8z^3b(\chi)-4z^2c(\chi)+4z\cos(\chi)+1}
\end{equation}
with, $I^{0}_{pump}= we^{2}\sin(\phi)z^{}_{p}\chi_{p}/\pi,
a(\chi)=2+2\sin(\chi)-\sin^{2}(\chi),
b(\chi)=cos^{2}(\chi)(\sin(\chi)-\cos(\chi)+1),
c(\chi)=\sin(\chi)+2\cos(\chi)-3\cos^{2}(\chi).$

\begin{figure}
\vskip 0.5cm
\protect\centerline{\epsfxsize=3.0in\epsfbox{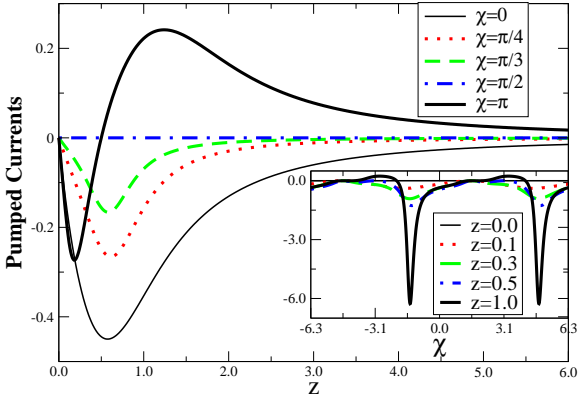}}
\caption{(Color online) The pumped current into the normal metal
lead at the interface between a normal metal and a charge density
wave interface  is of course finite in the tunnelling regime ($E
\ll \Delta$). In the main panel the pumped currents are as
function of the barrier strength $z$ for different values of the
phase difference $\chi$, while in the inset the currents are
plotted as function of the phase difference $\chi$ for different
values of the barrier strength $z$.}
\end{figure}

 The experimental realization of our structure wont be difficult.
Mesoscopic charge density wave interfaces have been around for
quite awhile now\cite{visscher1}. A metallic gate electrode
subject to an oscillating gate voltage is placed on top of the
charge density wave material, this arrangement can be effectively
used to modulate the phase of the charge density
wave\cite{visscher_ratchet}. Of-course a very similar structure to
that which is envisaged here has been experimentally realized by
Adelman, et. al., in Ref.\cite{adelman}. In the experiment of
Adelman, et. al., electric field induced variations of the charge
density wave order parameter lead to modulation of the
conductance. Further to modulate the interface delta function
barrier one can apply an oscillating voltage at the interface. The
experimental viability of this structure is of course guaranteed
since such type of make-up was theoretically envisaged to provide
for a charge density wave ratchet. The only difference will be
quantum interference effects dominating and the time dependent
voltages being in the adiabatic regime, i.e., at very low
temperatures and the system being in the mesoscopic regime.

 Finally to conclude this section it should be noted that
these three examples may not be unique there might be many other
examples of the distinctive nature of the rectified and pumped
currents which can be easily and unambiguously detected in
experiments.

\section{Conclusions}
\label{sec:conclusions} To conclude we have provided three
examples in which the pumped and rectified currents are so very
distinct. These examples provide an alternative and perhaps better
way to distinguish the rectified and pumped currents since these
go beyond looking just at the magnetic field symmetry of the
currents. The distinctive properties of the rectified and pumped
currents will also breakdown if the mesoscopic scatterer has
distinct spatial symmetries. In that case looking at the magnetic
field symmetry of the currents wont provide the solution. In the
first example given above the rectified currents are completely
unpolarized while the pumped currents are pure spin polarized, in
the second example we have net zero rectified currents for equal
strengths of the potential barriers while in example three the
rectified currents do not exist at all in the tunnelling regime
while the system pumps a definite amount of current both in to the
charge density wave material and the normal metal lead.

\section{Acknowledgments}
The author would like to thank Prof. G. E. W. Bauer for kindly
sending reference \cite{visscher1}. The author also thanks the
organizers of the workshop on Quantum information and decoherence
at Benasque, Spain from June 26-July 15 for providing funds to
attend. A major portion of the work was completed there.

\end{document}